# The Canada-France Redshift Survey V:
# Global Properties of the Sample

David Crampton[1]

Dominion Astrophysical Observatory, National Research Council of Canada, Victoria, Canada

O. Le Fèvre[1]

DAEC, Observatoire de Paris-Meudon, 92195 Meudon, France

S.J. Lilly[1]

Department of Astronomy, University of Toronto, Toronto, Canada

F. Hammer[1]

DAEC, Observatoire de Paris-Meudon, 92195 Meudon, France

## ABSTRACT

The Canada-France Redshift Survey is an unprecedentedly large sample of spectra of very faint $17.5 \leq I_{AB} \leq 22.5$ objects in 5 separate fields in which 85% of the target objects are securely identified. The photometric and spectroscopic data discussed in earlier CFRS papers are combined in this paper, and analyses are carried out to verify the integrity of the sample so that it can be confidently used for future scientific investigations.

The redshift histogram of the sample is presented for 591 field galaxies with secure redshifts. The median redshift is $<z> = 0.56$, and the highest redshift observed is $z \sim 1.3$; 25 galaxies have measured redshifts $> 1$. The distributions of magnitudes and colors demonstrate that galaxies at these high redshifts have very similar colors to those observed locally. The survey thus represents a major improvement in our knowledge of field galaxies at large look-back times.

Only $\sim 1$% of galaxies with $17.5 \leq I_{AB} \leq 22.5$ are as compact as stars (on images with FWHM $\sim 0''\!.9$) and comparison of the photometric and spectroscopic data show that only one galaxy was initially incorrectly classified spectroscopically as a star, and only two stars were misclassified as galaxies. It is demonstrated that the redshift distributions in the five fields are statistically consistent with each other, once the reduction in the effective number of independent galaxies due to small-scale clustering in redshift is taken into account.

The photometric properties of the spectroscopically-unidentified objects (15% of the sample) indicate that most are likely to be galaxies rather than stars. At least

---

half of these must have the same redshift distribution as the identified galaxies, and a combination of magnitudes, colors and compactness of the remaining unidentified galaxies is used to predict their redshifts. The majority are probably ordinary galaxies at the high redshift end of our sample, including some quiescent galaxies at z > 1.0, rather than some new or unusual population.

*Subject headings:* galaxies: evolution — galaxies: distances and redshifts, surveys

## 1. Introduction

The overall goal of the CFRS survey is to provide a large well-defined sample of field galaxies at high redshift (median z $\sim$ 0.6) for comparison with samples of the local galaxy population. An overview of the CFRS project and descriptions of the basic data gathered for the survey have been described in detail in the first four papers of this series (Lilly et al. 1995a; CFRS I, Le Fèvre et al. 1995a; CFRS II, Lilly et al. 1995b; CFRS III, and Hammer et al. 1995; CFRS IV).

In this paper, all of the spectroscopic, photometric and morphological data are brought together in order to understand the properties and limitations of the overall final galaxy sample and to establish a firm basis for future scientific investigations. Issues such as the reliability and completeness of the catalog are examined through intercomparison of a variety of photometric and morphological parameters. The data for three objects (out of over 1000) which were initially misclassified are corrected, and the overall statistical properties (magnitudes, colors, sizes) of the resulting complete sample of 591 field galaxies as a function of redshift are presented. The properties of the unidentified objects in the sample are investigated, and it is demonstrated that approximately half are basically identical to the identified sample. A "redshift predictor" is developed which suggests that many of the remaining unidentified galaxies are at higher redshifts where completeness problems were anticipated. Hence, it is possible to estimate, in a statistical sense, the properties of a large fraction of the unidentified objects and to demonstrate that our sample of 591 galaxies is representative of the population at z $\sim$0.6, corresponding to a look-back time of half the age of the Universe for $\Omega \sim 1$.

### 1.1. Features of the CFRS

Many features of the CFRS described in previous papers were designed to ensure the reliability and statistical integrity of the final data set. The more notable features are reviewed here.

Spectroscopic targets were selected in a completely unbiased way (e.g., without regard to compactness, morphology, colors, etc.) from a catalog of $17.5 \leq I_{AB} \leq 22.5$ objects based on deep I images of fields in five different directions. Possible surface brightness effects in the original photometric catalog were insignificant at the limit of our spectroscopic survey (CFRS I) and the spectroscopically identified subset are unbiassed in surface brightness relative to the photometric catalog (CFRS IV).

As described in CFRS II, The CFHT MOS spectrograph was used to obtain spectra with wavelength range 4250 - 8500Å of ∼80 objects simultaneously, arranged in three strips per slit mask. Every spectrum was reduced, and the corresponding redshift was determined, by at least three members of the CFRS team, again without knowledge of any other properties of the object. The final redshift and an estimate of its reliability was assigned by consensus. This procedure not only improved our confidence in the final redshift catalog, but also improved the completeness of the sample since one of the extractions of the spectra of these very faint galaxies was sometimes better than another. Well-defined criteria were employed to reject spectra for instrumental or technical reasons (e.g., bad CCD columns, overlapping zero orders, etc.) so that, for example, the success rate of determining redshifts of objects which were bright or which had strong emission-line spectra should not be favored by technical factors. All of the rejected objects were relegated to a 'supplemental catalog'.

A substantial fraction of the objects for which no reliable redshifts were initially determined were subsequently re-observed (CFRS III). These re-observations allowed us to empirically verify our reliability assessment which was initially based entirely on the appearance of the spectra, and to make strong predictions about the nature of the remaining unidentified objects, thereby effectively reducing the fraction of the sample whose nature remains truly unknown.

## 2. Summary of Basic Data

### 2.1. Photometric and Morphological Parameters

The selection of targets for the spectroscopic observations was based on isophotal $I$ magnitudes ($17.5 \leq I_{AB} \leq 22.5$) measured from very deep images to avoid any possible surface brightness selection effects. $B, V, I$ and $K$ $3''$–aperture magnitudes were also measured for most of the spectroscopic targets (details of the photometry are given in CFRS I). $V$ photometry is available for virtually all objects, and $B$ images were obtained for three of the fields. Due to the physical size of infrared detectors, the $K'$ imaging was more limited, but photometry was obtained for $\sim 80\%$ of the spectroscopic sample. In addition to magnitudes and colors, the isophotal radius, ellipticity and a compactness parameter Q, designed to distinquish stars from galaxies, were also measured for each object. As described in CFRS I, the Q parameter was normalized to the observed seeing on each image so that stars produce Q = 1 and galaxies yield increasingly larger values. As will be shown below, Q = 1.3 separates the majority of stars and galaxies.

### 2.2. Spectroscopic Parameters

As described in CFRS II, each redshift was assigned a confidence class based largely on the appearance of the spectra and our (combined) assessment of its reliability. As demonstrated in CFRS III, our initial estimates can be substantiated through an analysis of the repeat observations, and our confidence estimates were extremely good. To recap, the confidence classification for galaxies is as follows:

- Class 0: Redshift completely unknown.
- Class 1: Tentative redshift determined; estimated confidence $\sim 50\%$
- Class 2: Good redshift determined; confidence level $\sim 80\%$
- Class 3: Very good redshift determination; confidence level $> 95\%$
- Class 4: 100% certain redshift supported by many features.
- Class 8: Redshift based on one emission line identified as [OII] 3727 on the basis of the continuum shape (see CFRS III)
- Class 9: Redshift based on one emission line, tentatively identified with [OII] 3727.

A parallel notation was established for quasars by adding 10 to the above notations. Stars are simply assigned z = 0 with the same confidence classification from 1 to 4 as galaxies. Objects of any kind for which the redshift is unknown or not reliably determined, classes 0 and 1, are termed 'unidentified objects' in the following discussions.

The features present in our spectra upon which the redshifts are based were also recorded for all objects, as well as the equivalent width of [OII] 3727. Since the wavelength range recorded for each object was $4250 - 8500$Å, the strongest spectral features in galaxies at $z > 0.9$ may not have been observed and hence it is anticipated that our completeness in that redshift regime will be affected. Furthermore, the lack of spectral coverage below 4200Å combined with the declining response of the (thick) CCD in the blue, means that the important Ca II H&K lines and the 4000Å break are not available for identification of stars and very low redshift galaxies. This is particularly important for distinguishing the spectra of K stars and early-type galaxies with $z \sim 0.3$, since without the blue part of the spectrum, the strong Mg I 5180Å feature can be mistaken for the 4000Å break, particularly if an overlapping zero order image contaminates part of the spectrum. More details of these effects were discussed in CFRS IV. Objects which had to be discarded for purely technical reasons (see CFRS II) were relegated to a 'Supplementary Catalog' so that their redshifts are still available for analyses where their possible biases are irrelevant.

## 3. The Consistency of Photometric and Spectroscopic Parameters

All stages of the CFRS project were designed in such a way as to avoid, as far as possible, sources of bias. The selection of spectroscopic targets was done solely on the basis of magnitude, and the spectroscopic identification and classification were carried out without any knowledge of any other parameters of the objects (for details, see CFRS II). Thus we can use the photometric properties (colors and compactness) to check for consistency with the spectroscopic identifications. This is particularly relevant for objects with low quality spectra. We can also examine whether there is any substantial population of very compact galaxies, perhaps at high redshifts. The photometric imaging data for objects with good spectroscopic identifications is also used to establish the reliability of various techniques for separating galaxies from stars.

### 3.1. Compactness parameter vs. spectroscopic classification

A plot of the Q compactness parameter as a function of magnitude for spectroscopically-classified objects is shown in Figure 1. The lower panel (Figure 1a) shows spectroscopically-classified stars (open symbols) and galaxies (closed symbols) with confidence classes $\geq 2$. Figure 1a demonstrates that, as anticipated, virtually all of these high confidence class objects have almost certainly been correctly identified spectroscopically. Furthermore, the separation between stars and galaxies on the basis of compactness (for $\sim 0\rlap.{''}9$ seeing) appears to be quite good, although it decreases in reliability at magnitudes fainter than $I_{AB} \sim 20.5$ as its effectiveness depends strongly on signal-to-noise ratio. On the basis of these results, we empirically divide the diagram into three areas as shown:

- Region A: The small Q parameter indicates a point source.

- Region B: $Q > 1.3$ indicates that the object is extended (i.e., non-stellar).

- Region C: Objects in this area could be either point sources or extended.

The images and spectra of the objects which were apparently misclassified according to Figure 1a (i.e., "extended stars" in region B, or point-source "galaxies" in region A) were individually examined to determine whether any real misidentifications had occurred. The vast majority of the "extended stars" were simply the result of incorrect compactness parameters due to close companions. There are some truly compact galaxies; seven are indistinguishable from point sources. As will be discussed below, only three objects out of the whole sample of 1010 objects were discovered which appear to be spectroscopically misidentified when all spectroscopic and photometric data are considered. Comparison of Figures 1a and 1b demonstrates that most of the unidentified objects are likely to be galaxies rather than stars. The nature of the unidentified objects will be discussed in detail in section 6.

### 3.2. Colors vs. spectroscopic classification

The $(V - I)_{AB}$ and $(B - I)_{AB}$ colors are plotted versus the $(I - K)_{AB}$ colors for all spectroscopically observed objects in Figure 2. The two panels on the left (Figure 2a) are for all objects with confidence class $\geq 2$, and those on the right are for the 'unidentified objects' (class $\leq 1$). On average, there is good separation between stars and galaxies in the two-color planes, and the two dashed curves in each diagram are empirical curves drawn to delineate the approximate boundaries between stars and galaxies. Objects to the left (region S) are mostly stars, objects to the right in region G are predominantly galaxies, and objects in between are indeterminate (region I). Once again, the images and spectra of all objects with good spectroscopic classification which were deviant in either of the two-color diagrams– i.e., "spectroscopic galaxies" lying in region S, and any "spectroscopic stars" lying in region G in either color diagram were scrutinized for possible classification errors. Only two objects were noted that appear to have been spectroscopically misclassified; two of the same objects that were misclassified according to their compactness (the third object identified in §3.1 by compactness criteria is in the "indeterminate" area in the color-color diagram). The spectroscopic classifications of all other galaxies with discrepant colors are, without any doubt, correct since the spectra are unambiguously extragalactic. In some cases the colors were obviously suspect due to the faintness of the images or to crowding with nearby objects, but it appears that most of the unusual colors were due to unusual objects: 18 of the 29 galaxies with deviant colors displayed strong emission-line spectra.

### 3.3. Summary: misclassified objects

Thus, examination of objects with discrepant compactness and/or colors reveals that only three objects were initially misclassified on the basis of their spectra: CFRS 00.1608 is not a star but a galaxy with z = 0.269; CFRS 14.0664 and CFRS 14.0823 are both K stars, not galaxies at z ~0.3. These are classic aliases in redshift surveys of this type, due largely to the lack of blue response in the CCD detectors as mentioned above. In fact, one of us (SJL) noted that CFRS 14.0664 and 14.0823 were probably stars rather than galaxies during an overall error check of the spectra. This, plus the fact that there are only three of these errors out of over 1000 spectra, led us to go back and correct our basic spectroscopic catalog (and the corresponding tables in CFRS II and CFRS III) rather than to keep these objects separate.

### 3.4. Confidence of Photometric Star-Galaxy Separation Techniques

The analyses above demonstrate that most stars can be distinguished from galaxies on the basis of compactness (with good seeing!) and colors. However, there are a number of definite galaxies that lie on the stellar loci in either the compactness or color-color diagrams or both. Since our target selection was completely unbiased with respect to color or compactness, it is possible to estimate the fraction of galaxies which would be missed if targets had been selected on the basis of these parameters.

Of the 701 objects with spectroscopic confidence classes ≥3, 7 galaxies (and 4 quasars) lie in region A of Figure 1a, i.e., they were indistinguishable from point sources. The remaining two quasars have detectable host galaxies and are "non-stellar". An additional 19 galaxies lie in region C, the region at the faint limit where the distinction between stars and galaxies is more difficult. Hence if all objects lying in region A had been rejected, and all in B and C observed, only 1% of the galaxies would have been lost (and 153 or 86% of the stars would have been rejected). Nevertheless, inclusion of all objects in our initial target list allows us to unequivocally state that only ∼1% of galaxies with $17.5 \leq I_{AB} < 22.5$ are as compact as stars on images with subarcsecond seeing (FWHM ∼ 0″.9).

The situation with regard to the confidence of star-galaxy separation by color selection is worse. Figure 2 illustrates that many galaxies have quite blue $(I - K)$ colors and are indistinguishable from stars. For example, among the 29 galaxies (6.7% of the total) which lie in region S in the $(V - I) - (I - K)$ diagram, 18 have strong emission-line spectra. Thus, selection on the basis of $VIK$ colors would give a strong bias against this type of object. Selection based on the $(B - I) - (I - K)$ colors appears to be more efficient: only 5 of 182 galaxies (2.7%) would have been incorrectly classified as almost certain stars from their colors. But again, galaxies with strong emission-lines would be discriminated against.

## 4. Global properties of the CFRS sample

A summary of the CFRS catalog is given in Table 1. The catalog consists of 1010 objects, of which 67 are in the supplementary catalog, yielding a statistically complete sample of 943 objects. Taking into account the very faint limits of this survey (many objects have B > 24), a very high fraction (85%) of the objects have been identified, and most of these have very secure redshifts. A summary showing the fraction of the confidence classes for all objects in the complete catalog is given in Table 2. Note that 70% of all classifications have confidence classes 3, 4 or 8, – and the empirical tests discussed in CFRS III indicate that these identifications are correct at the 95% - 100% level.

### 4.1. Distribution in redshift

The 591 galaxies with secure redshifts in the sample have a median redshift of <z>∼ 0.56, and hence they comprise by far the largest sample of high redshift field galaxies yet assembled. The redshift distribution of the identified galaxy sample as a whole is shown in Figure 3. The open rectangles show the relative areas occupied by stars, unidentified objects and quasars. This figure demonstrates that we have succeeded in our goal of obtaining a large sample of galaxies with z ∼0.6, corresponding to a look-back time of half the age of the Universe for $\Omega \sim 1$. Apart from the I-band samples published by Lilly (1993) and Tresse et al. (1993) which led up to the CFRS survey, the most comparable samples are the B-band selected sample of Glazebrook at al. (1995) and the K-band selected sample of Songaila et al. (1994). These each contain less than 50 field galaxies with z > 0.5 as compared with 350 in the CFRS sample. Furthermore, the I-band selection means that high redshift subsamples of the survey can be compared directly with those at lower redshift, thereby avoiding the problems associated with the very different selection criteria encountered when local samples are compared to those at high redshifts.

This survey is also the first to contain a sizable number of field galaxies at z ∼ 1. There are 25 with z > 1, and as will be discussed in §6.5, there is considerable evidence that several more, particularly early-type, galaxies with z > 1 must be present among our unidentified objects.

### 4.2. Color and magnitude as a function of redshift

Redshifts for all identified galaxies (confidence class ≥2) are shown plotted as a function of $I_{AB}$ magnitude in Fig. 4. For reference, the tracks show the redshifts for non-evolving galaxies with $M_{AB}(B)$ = -21.0 and three spectral energy distributions (E, Sbc, Irr) from Coleman, Wu and Weedman (1980). These converge at $z \sim 0.9$ because at this point the observed I-band matches the rest B-band.

Figure 5 shows the $(V - I)_{AB}$ colors plotted versus the redshift for the identified galaxies, again with curves showing colors for the same three unevolving spectral energy distributions from Coleman, Wu and Weedman (1980). The figure demonstrates that the majority of the galaxies apparently have perfectly normal colors, even at large look-back times. The full range of galaxy types, from the reddest to the bluest, is observed at all redshifts up to at least z ∼ 0.9. One obvious feature in our data at the highest redshifts is that very few galaxies redder than Sbc's are included at z > 0.9, whereas bluer galaxies are observed up to z = 1.3. This almost certainly arises from difficulties in securing identifications for absorption line objects at $z > 1.0$ in spectra limited to wavelengths shortward of 8500 /AA . This selection bias is discussed more fully in CFRS III and CFRS IV.

The loci of redshifted colors as a function of redshift, again calculated for different galaxy spectral energy distributions from Coleman, Wu and Weedman (1980), are shown in Figure 6a (left panels), superimposed on the observed $(V - I)_{AB}$ vs $(I - K)_{AB}$ colors of the galaxies with redshift classes ≥2. The equivalent diagram for galaxies without secure redshifts is shown to the right in Figure 6b. The four dashed lines indicate different fiducial redshifts increasing from z = 0 on the left, to z = 2 on the right. The three solid lines indicate three different galaxy spectral energy distributions from Irr on the bottom to E at the top. Although there is an ambiguity at low z, it is obvious that many of the unidentified objects are in the "high redshift early-type galaxy" area. In fact, 21% of the unidentified objects in the $(V - I)_{AB}$ - $(I - K)_{AB}$ diagram are in the "z > 1, redder-than-Sbc" area. This again supports our hypothesis that the limitations of our observing technique has prevented us from identifying a number of the highest redshift objects in our sample, particularly those without an [OII] 3727 emission line. The probable fraction of $I_{AB} \leq 22.5$ galaxies at z > 1 will be discussed further in §6.5.

### 4.3. Completeness as a function of magnitude

The box at the right side of Figure 3 represents the area occupied by the unidentified objects in our sample. Overall, the redshift identification success rate is 85%, and it is obvious that the distribution of redshifts of $17.5 \leq I_{AB} \leq 22.5$ galaxies cannot be very different from that shown in Figure 3 unless the unidentified objects have a completely different distribution. In fact, as will be shown in §6.2, it can be demonstrated that the redshift distribution for at least half of the unknowns is likely to be the *same* as for the identified sample and, based on their colors, magnitudes and size, the redshifts of the remaining 7% of the objects in the CFRS sample cannot be very different from those of the identified fraction. We conclude that the redshift distribution shown in Figure 3 is thus truly representative of galaxies with $17.5 \leq I_{AB} \leq 22.5$.

As expected, the identification rate decreases as a function of magnitude. Figure 7 shows the identified fraction of all objects (solid line) and galaxies (dashed line) as a function of magnitude. The overall success rate (85%) compares very favorably with those of other large redshift surveys, even those with a considerably brighter magnitude limit. As mentioned above, this is partly attributable to the fact that three independent reductions and identifications were made of all spectra.

### 4.4. Quasars

Only 6 quasars, listed in Table 3, were discovered during the course of the CFRS survey. They do not appear to be grossly different from those detected in other surveys based on e.g., colors, emission lines, radio properties, etc. Since our survey is not biased in any way against discovery of quasars, these six form a complete sample to a relatively faint magnitude. Further discussion of this sample will be given by Schade et al. 1995.

### 5. Statistical significance of field-to-field variations

With so many galaxies in our sample, the "non-smooth" nature of the redshift histogram in Figure 3 is perhaps initially surprising, e.g., the "peak" at z ∼ 0.2 or the "hole" at z ∼ 0.4. Various internal tests of the data were carried out, as described in CFRS IV, to ensure that any biases for or against particular values of redshifts due to technical effects are insignificant, so we believe the shape of the histogram is real. We were especially concerned that our observational technique (which yields three tiers of spectra) might introduce some bias since the overlapping zero order spectra obliterate a part of the spectra below. Both visual examinations of the spatial distribution of the unidentified objects, and the objective test described in CFRS IV showed no evidence of such an effect. All analyses indicate that the redshift distributions are not affected by instrumental effects or by our data reduction techniques. Rather, there is good evidence that the detailed shape of the redshift histogram arises from the statistics of the "picket fence" structures which were encountered in our 5 lines-of-sight.

The redshift histograms of the 5 fields are shown individually in Figure 8. Even though the differences among the fields appear striking, the statistical significance of the apparent large-scale structures in the redshift distributions is low. The main reason for this is that galaxies are correlated on small scales (see Le Fèvre et al. 1995b), so that whenever a galaxy is found in a structure, other galaxies will be found nearby and the visual impression of large-scale structures is strengthened. Thus, the number of *independent* galaxies and the corresponding statistical significance of potential field-to-field differences are substantially reduced.

In order to determine the significance of the differences in the redshift histograms among our five fields, we initially compare the individual fields to the sample as a whole. A "predicted" redshift distribution, was constructed from the luminosity function (Lilly et al. 1995c; CFRS VI) – all of the "features" in the original distribution (Figure 3) are smoothed out, as expected. The redshift distribution for each field is then compared to this smooth distribution using both a chi-squared test with 0.1 redshift binning, and a Kolmogorov-Smirnov (K-S) test with 0.01 redshift binning. On the face of it, the results for each field (see Table 4) indicate that the redshift distributions for all fields are apparently inconsistent with the smooth distribution of the CFRS sample as a whole. However, this is without any correction for small-scale galaxy-galaxy correlation.

The correction for small-scale clustering in each field was quantified by calculating the average number of galaxies, $N_c$, associated with each galaxy (including itself) above the smoothed redshift distribution. The average $N_c$ is not strongly dependent on redshift and for each field varies between 1.7 and 3.0 as shown in the second column of Table 5, being highest in the fields with most galaxies since, as the background goes up, the number of associated galaxies also goes up. When these numbers are applied as a correction factor to determine the "effective number" of independent galaxies, $N_{eff}$, in each field, the chi-squared and K-S tests show (Table 5) that the differences between the individual fields and the whole sample become insignificant. In both tests the most significant difference is for the $00^h$ field, as one might expect from the appearance of the histograms in Figure 8, but it also has the smallest number of galaxies.

A straightforward intercomparison of the fields among themselves also demonstrates that there are no significant differences in their redshift distributions. When corrected for the small-scale galaxy-galaxy correlation, K-S tests show that the probability that $00^h$ and $10^h$ field distributions are different is only p$\sim$ 0.12, and for the $14^h$ and $22^h$ fields, it is p$\sim$ 0.07, i.e., once again the significance is very low.

In summary, we believe that the large scale field-to-field variations that appear to be present in the redshift distributions shown in Figure 8 are simply artefacts produced by the small-scale correlations among galaxies. The latter correlations will be examined in more detail by Le Fèvre et al. (1995b; CFRS VIII).

## 6. The Unidentified Objects

### 6.1. Are the unidentified objects stars or galaxies?

Figures 1 and 2 can be used to estimate the properties of the unidentified objects, those with spectroscopic classifications 0 and 1. The upper panel of Figure 1 (Fig. 1b) clearly shows that the majority of the objects are extended and probably are galaxies. Only 5 objects lie in area A (most probably stars), and an additional 14 lie in area C (indeterminate). The right panels of Fig. 2 show that the majority of objects for which we have K photometry also have the colors of galaxies. To combine the compactness and color information, a photometric classifier was devised, such that if both the colors and compactness indicate that an object is a star, a class = -2 was assigned; if either gave an indication of a star, a class = -1 was assigned; and so on, until if both signify that an object is a galaxy a number = +2 was assigned. Class = 0 objects include both the cases where the indicators conflicted, and the objects for which the data were not available. In other words, the grades -2 to +2 represent increasing degrees of certainty that an object is a galaxy on the basis of photometric parameters.

A histogram of these classes for the 146 unidentified objects is shown in Figure 9. For only two objects do both the colors and compactness indicate that the object is a star, and in one of these cases it was also spectroscopically classified as a star, albeit at a low confidence level. Given the fact that Figure 2, in particular, shows that galaxies can scatter onto the stellar loci, it is problematic how to decide whether the remaining 17 compact ($Q < 1.3$) objects are stars or galaxies. The upper panel of Figure 1 shows that roughly one-third (i.e., 6) of the unidentified compact objects (region C) are likely to be stars. Even though a few confirmed stars lie above the $Q > 1.3$ line, it is probable that most are galaxies since in general it is evidently spectroscopically easier to identify stars than galaxies at faint limits. We thus accept the 7 class -1 and -2 objects as stars and adopt the remaining 139 objects as galaxies. Although it is obvious that we cannot be certain that these 7 objects are stars, they should be representative of the sample as a whole. The "presumed stars" are: CFRS 00.1471, 03.1091, 03.1128, 10.1212, 10.2425, 14.0760, and 22.0272.

### 6.2. Analysis of repeat observations

An estimate of the proportion of unidentified galaxies which have a "normal" redshift distribution (i.e., the same as the whole sample) can be gained from repeated observations of the same objects during our survey. As discussed in CFRS III, of 99 spectroscopically-unidentified (initially) objects that were re-observed (with identical techniques, exposure times, etc.), a high fraction, 70%, were subsequently identified and only 29 were "persistent failures". Since some of these repeat observations were of targets for which the initial data were somewhat poorer than usual, we estimate that our recovery rate for the remaining 110 galaxies (139 unidentified galaxies minus these 29 persistent failures that are unlikely to be identified) would probably be somewhat less than this 70% recovery rate, say~ 60%. In other words, further observations would readily yield redshifts for ~65 (110 × 60%) of the objects still listed as unidentified. As will be shown in §6.4, the "recovered failures" have the same redshift distribution as the sample as a whole and hence, by extnsion, these 65 galaxies will also have the same redshift distribution as the sample as a whole. We thus have considerable confidence that the redshift distribution for almost half of the 139 unidentified galaxies is identical to that of the identified fraction. Additional evidence of this canbe derived from the properties of the unidentified objects, as shown in the following sections.

### 6.3. A Redshift Predictor

Figures 4, 5 and 6 demonstrate that, as expected, the magnitudes, and colors of target galaxies are correlated, albeit weakly, with redshift. In this section we attempt to combine these data, plus the size (compactness parameter Q) in order to obtain a better estimate of the redshift of the unidentified objects than could be obtained from any of the parameters on its own.

The imaging data for all of the galaxies in the complete sample with confidence class > 2 were used as a 'learning sample' to estimate the probable redshifts of the unidentified objects. A redshift parameter R was developed empirically from the redshifts of nearest neighbors in a multi-dimensional parameter space consisting of the $I_{AB}$ magnitude, the $(V - I)$ colors, the $(I - K)$ colors where available, and the Q parameter. Tests of a variety of different weighting schemes for the parameters, and trials involving different numbers of nearest neighbors were carried out to explore the parameter space. None of the parameters seemed particularly sensitive except that the Q compactness parameter had to be given a low weight relative to the others, as expected. The final parameter which was adopted involves weighting the above parameters in the ratio 4:4:8:1 respectively, and taking the median redshift of the three nearest neighbors as the predicted value. The results of applying this estimator to the for the galaxies with secure redshifts are shown in Figure 10a. Although the predictor gives discrepant results for some galaxies, it does quite well for the majority. The Spearman rank correlation coefficient is $r_s = 0.46$ for n = 589 (P >> 99.99%) and the r.m.s. difference between predicted and observed redshifts is 0.22. As a semi-independent test of the method, the predictor was applied to all "recovered failures" – objects which were initially spectroscopically unidentified but subsequently identified by repeat observation(s). A comparison of their predicted redshift from the estimator with that subsequently measured is shown in Figure 10b. Once again, the predictor apparently works well - the rank correlation coefficient is 0.50 (P > 99.99%) and the r.m.s. redshift error is 0.23. While we are not advocating use of this predictor to estimate redshifts of individual objects, it can probably be used to estimate the statistical properties of the unidentified objects in our sample. An obvious shortcoming of the estimator is that it cannot predict redshifts different from the sample from which it was constructed. The empirically determined accuracy of our predictor can not neccessarily be translated to that of other predictors (e.g., Loh and Spillar 1986) because the accuracy will depend on the choice of passbands, the photometric accuracy and the choice of templates.

### 6.4. Redshifts of galaxies with z < 1

Histograms showing the actual and predicted redshift distributions of the 'recovered unidentified objects' (the 70 galaxies discussed in §6.2 for which a subsequent observation yielded a redshift) are shown in Figures 11b and 11c respectively. The two distributions are not statistically significant different from each other, demonstrating the usefulness of the predictor. As noted in §6.2, the distribution of redshifts of the 'recovered failures' is statistically identical to that of the whole sample (Figure 11a), indicating that the initial failure was due to a technical problem rather than some completely different type of spectrum.

Figure 11d shows that the redshift distribution predicted for those objects for which repeated observations did **not** yield a redshift, differs considerably. According to the predictor, most of these are at z > 0.7, in agreement with our hypothesis that our sample is subject to incompleteness at the highest redshifts since many of the prominent spectral features are redshifted out of the observed wavelength range (i.e., to wavelengths > 8500Å). Figure 11e shows that the redshift distribution of all the 139 unidentified galaxies is higher than average, once again implying that the ∼50% of objects for which repeated observation would be unlikely to yield a redshift must be at higher redshifts (since we have shown in §6.2 that half of them must have a "normal" distribution).

Hence, the redshift estimator supports our hypothesis that the redshift distribution of about half of the unidentified objects can be confidently assumed to be the same as for the identified fraction, and predicts that the redshifts of the remaining half are likely to be higher than average, at z > 0.5. It should be remembered, however, that the redshift estimator cannot, by construction, predict redshifts greater than observed in the identified sample.

### 6.5. Redshifts > 1

Several lines of argument suggest that a number of unidentified objects in our sample may lie at z > 1. The luminosity function at 0.5 < z < 0.75, where our sample contains a substantial number of galaxies, and where the effects of the incompleteness in spectroscopic identification must be small, can be used to obtain a crude estimate of how many high redshift galaxies are missing (assuming there is no additional evolution). As shown in Figure 12, the luminosity function derived in CFRS VI for the sample with $0.5 < z < 0.75$, predicts that a total of 42 galaxies should have been observed at $z > 1$, 20 more than identified. On the two-color plots (Figure 6), we see about this number of unidentified red galaxies whose colors are consistent with $z > 1$. We therefore predict that at least 20 of the unidentified objects will be at $z > 1$, and they will probably be red, early-type galaxies. Given the consistency among the information from the colors, redshift predictor, and luminosity function, the number of unidentified galaxies likely to be at $z > 1$ is adopted to be 20.

### 6.6. Nature of the unidentified galaxies

In summary, of the 139 unidentified galaxies, our analyses show that 65 almost certainly have a redshift distribution similar to that of the sample as a whole, 20 are likely to be at $z > 1$, and the redshift estimator predicts that most of the remaining 54 are likely to be at $0.5 < z < 1.0$. Undoubtedly, many of the objects remain unidentified due to the increasing technical difficulties of deriving redshifts at $z > 0.7$ with our observational strategy (see also the discussion in Crampton et al. 1995). Apart from the putative red galaxies at $z > 1$ which are unrepresented in the spectroscopically identified objects, most of our 'unidentified objects' are thus quite normal, and do not represent some extreme, exotic population. Their properties appear to be very similar to the remainder of the sample, and hence we are able to investigate the nature of the galaxy population to $z \sim 1$ based on our sample with considerable confidence.

### 7. Summary

Internal tests indicate that the CFRS sample of 591 galaxies with secure redshifts is to a large degree free of biases, and hence provides the first large reliable sample of high redshift field galaxies.

The compactness and colors of all objects in the CFRS catalog were used to search for objects which might have been spectroscopically misclassified or have unusual colors. Detailed examination of all 1010 objects reveals that only two stars had been mistaken as galaxies on the basis of their spectra, and only one real galaxy had been mistaken for a star. Examination of the galaxy colors demonstrates that ~7% have indistinguishable $VIK$ colors, in our data, to stars, and more than half of these are strong-emission-line galaxies which might be excluded if target selection had been based on colors. Very few galaxies (only ~ 1% with seeing ~ $0\rlap{.}''9$) are as compact as stars, even at $z \geq 1$.

The median redshift of the galaxies with secure redshifts is $z \sim 0.56$, and the sample extends to at least $z \sim 1.3$ with 25 galaxies at $z > 1$. The colors of the galaxies span the full range observed locally, out to at least $z \sim 0.9$.

– 15 –Although field-to-field variations in the redshift distributions among our 5 fields are visually quite apparent, tests demonstrate that they are of marginal statistical significance, once the effects of small-scale galaxy-galaxy correlations have been taken into account. The small scale clustering acts to effectively reduce the number of independent galaxies in the sample.

The completeness of secure spectroscopic identification, $\sim 85\%$ for the sample as a whole, is high for this type of survey, particularly at such a faint magnitude limit. The morphology and colors of the unidentified objects indicates that 95% of them are galaxies. Furthermore, analysis of repeat observations shows that the redshift distribution for almost half of these must be the same as that of the spectroscopically identified galaxies, effectively reducing the spectroscopic incompleteness to $\sim 7\%$. Through comparison of the magnitudes, colors and sizes of the galaxies in this remaining unidentified fraction with those of the identified sample, it is shown that most are likely to be early-type galaxies at high redshifts where our observational techniques, and particularly the long wavelength cut-off of our spectra at 8500Å, made the securing of identifications difficult; it is estimated that 20 of the unidentified galaxies are likely to be at $z > 1$, and many of the remainder at $0.5 < z < 1$.

We conclude that the CFRS catalog represents a sound basis for future scientific investigations of the population of normal galaxies at large look-back times. Our studies of the evolving luminosity function, of the clustering correlation length and of the morphologies and spectra of these galaxies will be presented in future papers in this series.

We thank the CTAC and CFGT for their allocations of time for this relatively large project, and the directors of the CFHT for their continuing support and encouragement. The referee made a number of useful suggestions which improved the presentation of this paper. SJL's research is supported by the NSERC of Canada and travel support from NATO is gratefully acknowledged.## REFERENCES

Coleman, G.D., Wu, C.C., & Weedman, D.W., 1980, ApJS, 43, 393

Crampton, D., Morbey, C.L., Le Fèvre, O., Hammer, F., Tresse, L., Lilly, S.J., & Schade, D.J., 1995, in "Wide-Field Spectroscopy and the Distant Universe", Proc. 35th Herstmonceux Conf., ed. S. Maddox, World Scientific, in press

Le Fèvre, O., Crampton, D., Hammer, F., Lilly, S.J., and Tresse, L. 1994, ApJ, 423, L89

Le Fèvre, O., Crampton, D., Lilly, S., Hammer, F., Tresse, L., 1995a, (CFRS II)

Le Fèvre, O., Hammer, F., Lilly, S., Crampton, D.,1995b, (CFRS VIII)

Glazebrook, K., Ellis, R.S., Colless, M.M., Broadhurst, T.J., Allington-Smith, J.R., Tamvir, N.R., & Taylor, K., 1995, MNRAS, in press

Hammer, F., Crampton, D., Le Fèvre, O., Lilly, S.J., 1995 (CFRS IV)

Table 1: Composition of CFRS Sample

| Type | Number | Percent |
|---|---|---|
| Galaxies | 591 | 63 |
| Stars | 200 | 21 |
| QSO | 6 | 1 |
| Unidentified | 146 | 15 |
| Total | 943 | 100 |

Table 2: Fractions of objects in each confidence class

| Confidence class | N | Fraction |
|---|---|---|
| Class 4 | 330 | 35% |
| Class 3 | 313 | 33% |
| Class 8 | 29 | 3% |
| Class 9 | 33 | 3% |
| Class 2 | 92 | 10% |
| Class 1 | 53 | 6% |
| Class 0 | 93 | 10% |
| Sum | 943 | 100% |

Table 3: Quasars in the CFRS sample

| CFRS | z |
|---|---|
| 00.0207 | 1.352 |
| 03.0603 | 1.048 |
| 03.0106 | 2.07 |
| 14.0198 | 1.6034 |
| 14.1303 | 0.9850 |
| 14.1567 | 0.4787 |

Table 4: Significance of field-to-field variations

| Field | N | $\chi^2$ | p | K-S | p |
|---|---|---|---|---|---|
| 00 | 55 | 3.0 | <0.001 | 1.76 | 0.003 |
| 03 | 175 | 2.8 | 0.001 | 1.27 | 0.07 |
| 10 | 150 | 3.9 | <0.001 | 2.17 | <0.001 |
| 14 | 162 | 2.8 | 0.001 | 1.47 | 0.03 |
| 22 | 106 | 2.2 | 0.01 | 1.51 | 0.03 |
| ALL | 650 | 3.0 | <0.001 | 1.83 | 0.003 |

Table 5: Corrected significance of field-to-field variations

| Field | $N_C$ | $N_{eff}$ | $\chi^2$ | p | K-S | p |
|---|---|---|---|---|---|---|
| 00 | 1.76 | 31 | 1.7 | 0.06 | 1.33 | 0.06 |
| 03 | 2.91 | 60 | 0.9 | 0.55 | 0.74 | .... |
| 10 | 3.00 | 50 | 1.3 | 0.24 | 1.25 | 0.09 |
| 14 | 1.83 | 90 | 1.5 | 0.12 | 1.09 | 0.18 |
| 22 | 2.14 | 50 | 1.0 | 0.45 | 1.03 | .... |
| ALL | 2.33 | 282 | 1.3 | 0.24 | 1.20 | 0.11 |

Fig. 1.— (a) The compactness parameter Q plotted against $I_{AB}$ for all spectroscopically-classified objects with confidence classes $\geq 2$ (lower panel) and (b), for the unidentified objects (upper panel). The spectroscopic confidence classes associated with the various objects are shown in parentheses. The regions A (stars), B (extended objects) and C (indeterminate) are discussed in the text.

Fig. 2.— The $(V - I)_{AB}$ and $(B - I)_{AB}$ colors plotted versus the $(I - K)_{AB}$ colors for all spectroscopically-classified objects with confidence class $\geq 2$ (left), and for the 'unidentified objects' (right). The dashed lines delineate regions which contain mostly stars (S), mostly galaxies (G), and an intermediate region where the colors of stars and galaxies are indistinguishable (I).

Fig. 3.— The redshift distribution of all 591 galaxies in the CFRS statistically complete sample. The box on the left shows a representative area occupied by stars, that on the upper right represents the unidentified fraction of objects, and the small box shows the area occupied by quasars.

Fig. 4.— Hubble diagram for all 591 galaxies with secure redshifts in the CFRS sample. The dashed lines represent the redshifts of non-evolving galaxies with different spectral energy distributions and $M_{AB}(B) = -21.0$ (see text).

Fig. 5.— $(V - I)_{AB}$ colors of all 591 galaxies with secure redshifts in the CFRS sample, with tracks based on unevolving spectral energy distributions from Coleman, Wu and Weedman (1980).

Fig. 6.— The $(V - I)_{AB}$ and $(B - I)_{AB}$ colors plotted versus the $(I - K)_{AB}$ colors for all definite (confidence class $\geq 2$) galaxies in our sample (left), and for the galaxies without redshifts (right). Typical tracks (solid lines) are shown for three classes (Irr at bottom, Sbc in the middle, and E at top) of galaxies, with four fiducial redshifts (z = 0, 0.5, 1, 2) indicated by dashed lines.

Fig. 7.— Completeness of the redshift identification success rate as a function of magnitude for the whole sample (solid line) and for galaxies (dashed line).

Fig. 8.— Redshift distributions for the five individual fields. Although the variations among the distributions appear large, they are consistent with all being drawn from a common population.

Fig. 9.— A histogram of the numbers of spectroscopically-unidentified objects that satisfy various photometric color and compactness criteria. Both criteria indicate that objects with class = 2 are galaxies; both criteria indicate that objects with class = -2 are probably stars. In other words, as the class increases, there is increasing probability that an object is a galaxy based on color and compactness criteria.

Fig. 10.— Estimated redshifts from the redshift predictor discussed in the text compared to the measured redshifts for a) all 591 CFRS galaxies with secure redshifts and b) objects for which repeated observations yielded a redshift.

Fig. 11.— Redshift distributions (from top to bottom) of a) measured redshifts for the complete sample of 591 galaxies, b) galaxies for which redshifts were derived from subsequent observations, c) predicted redshifts for the latter galaxies, d) predicted redshifts for galaxies for which repeat observations did not yield a redshift, and e) all galaxies in the CFRS sample with unknown redshifts.

Fig. 12.— Predicted redshift distribution (solid line) from the luminosity function derived for galaxies with $0.5 < z < 0.75$ in CFRS VI, compared to the observed distribution (hatched area).

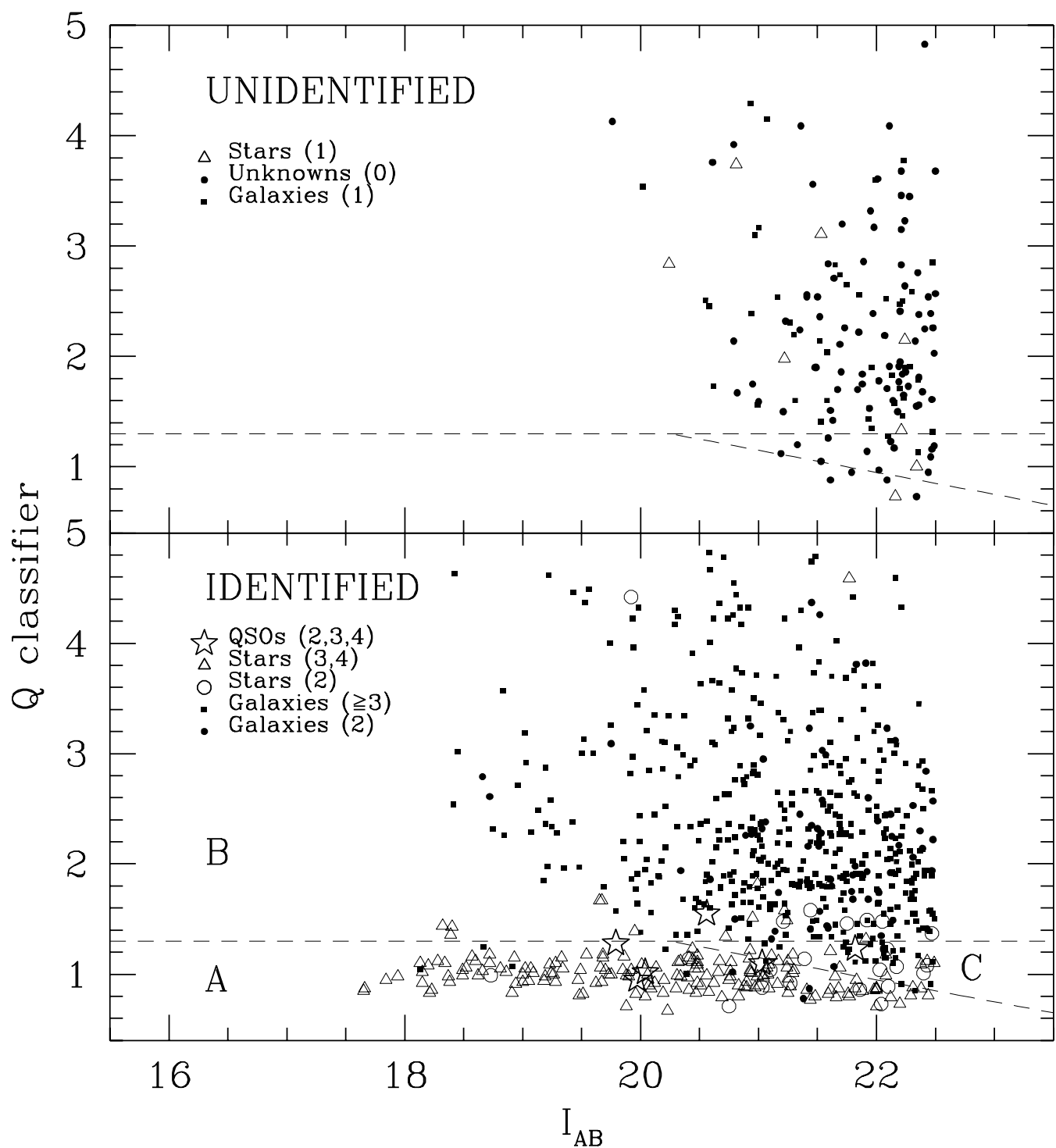

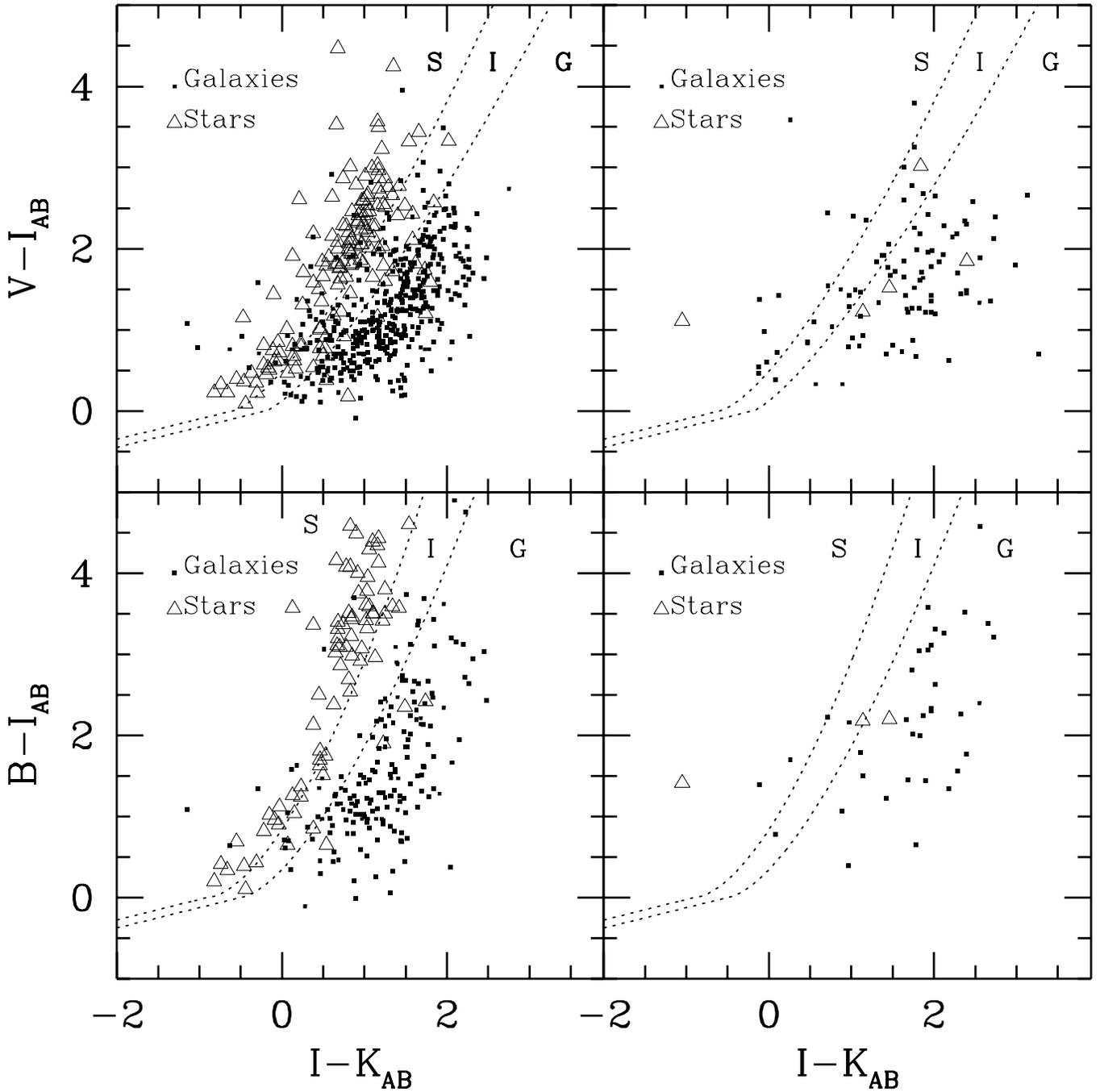

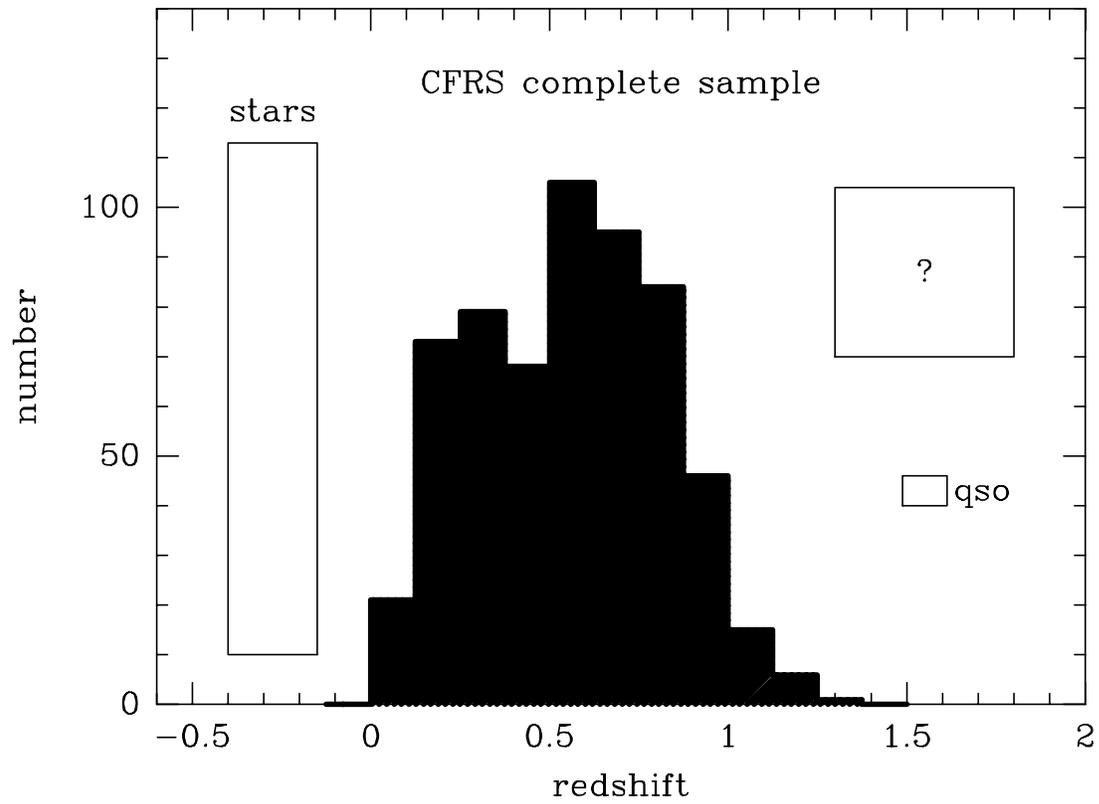

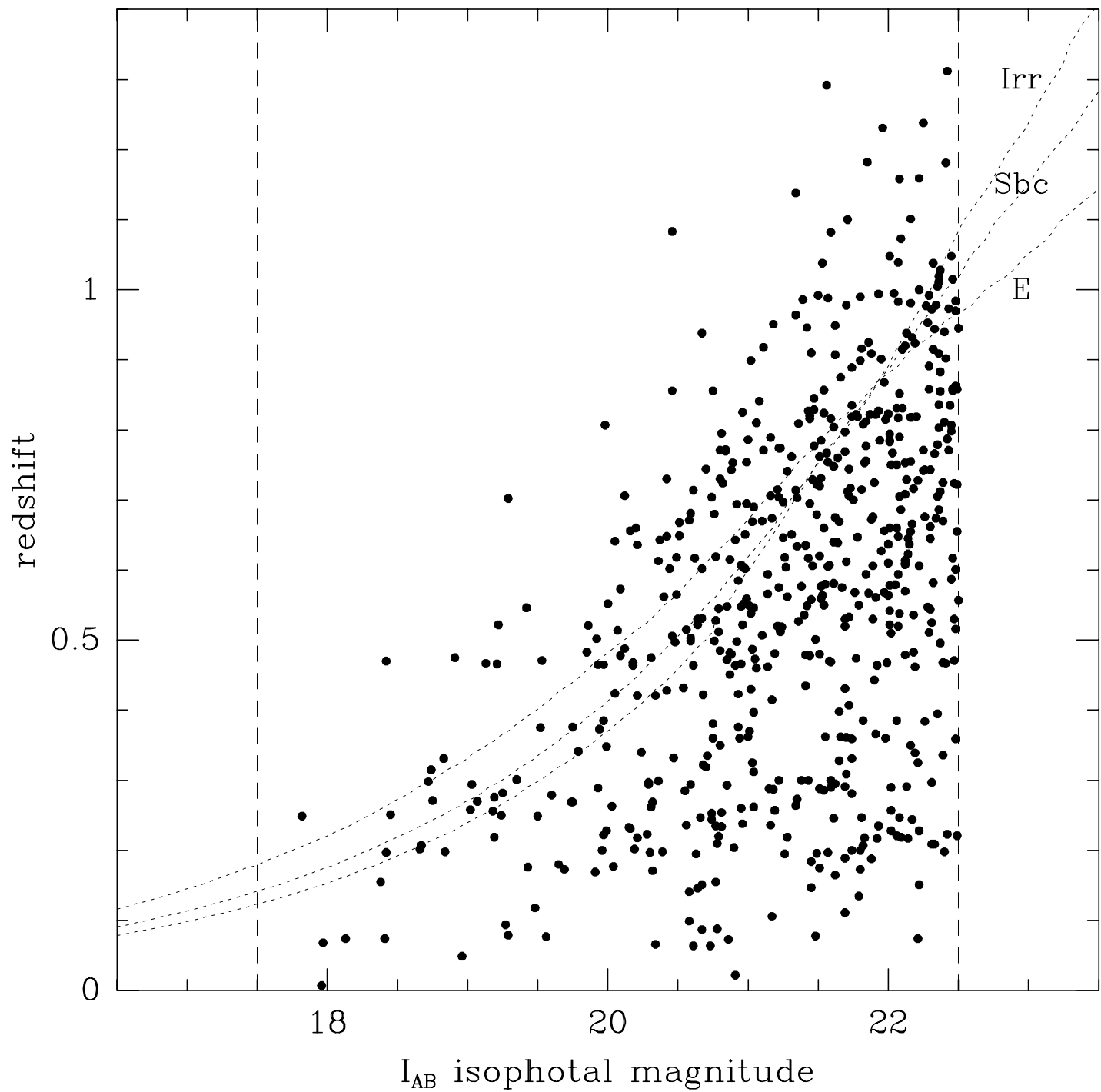

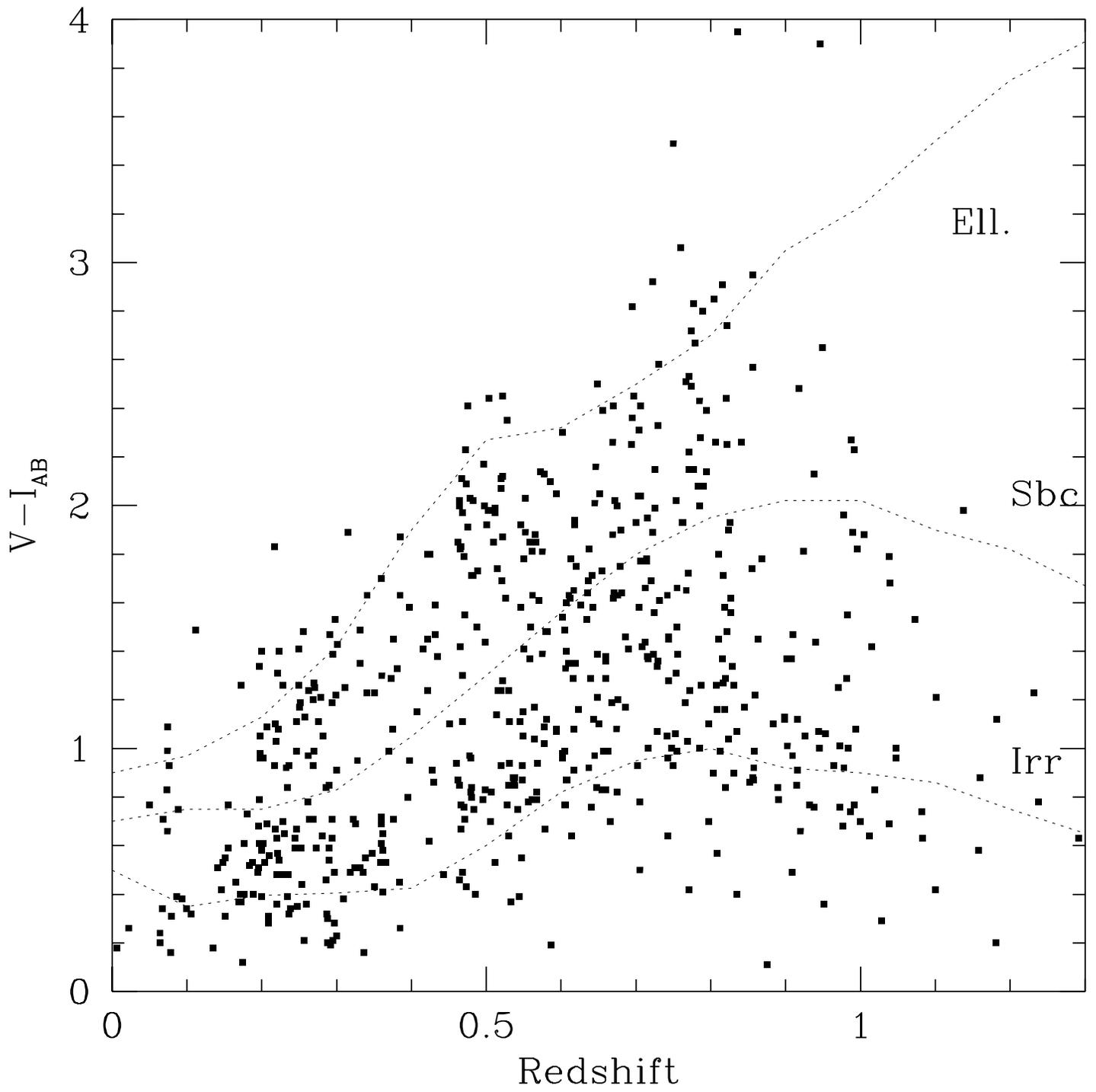

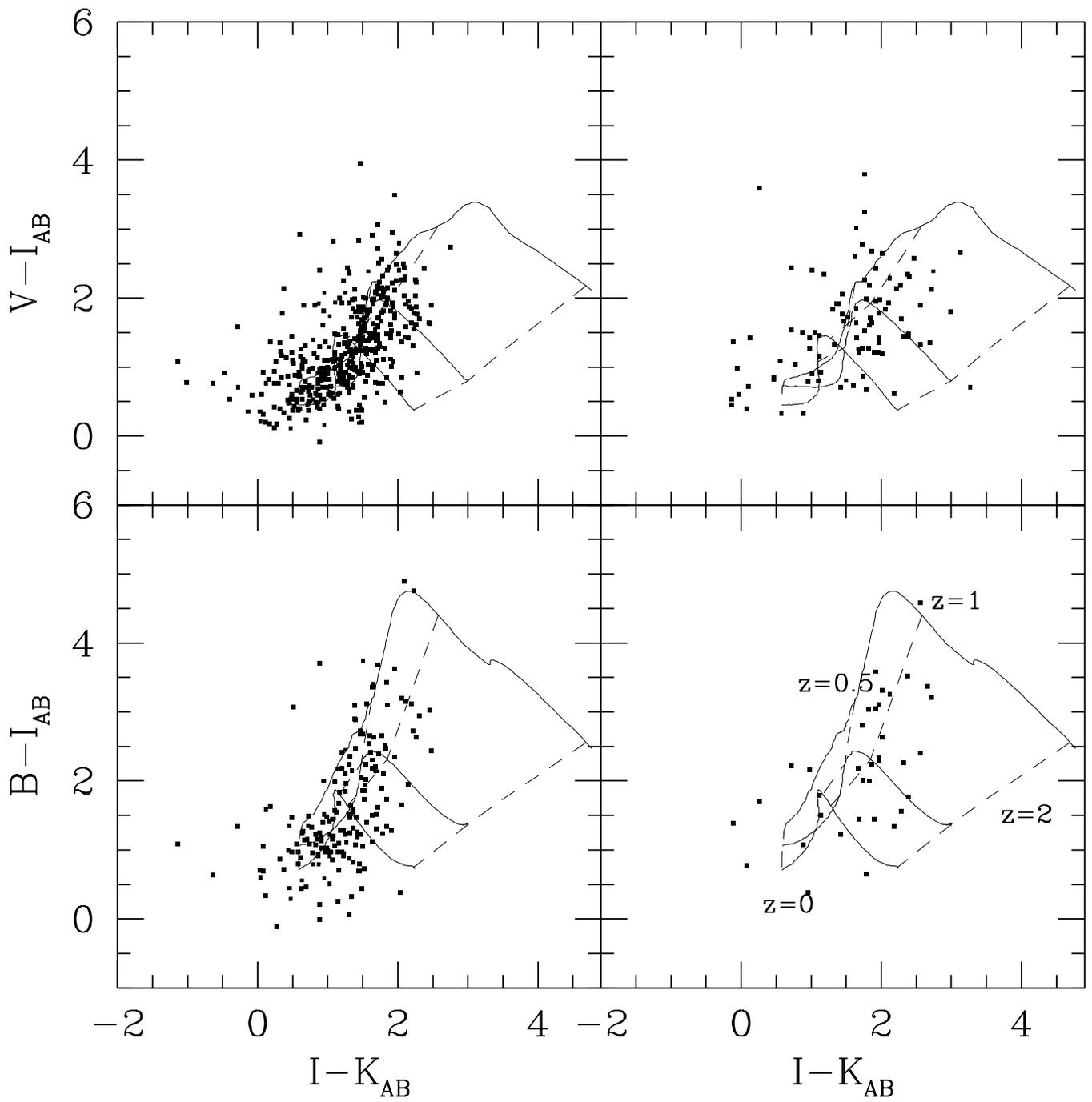

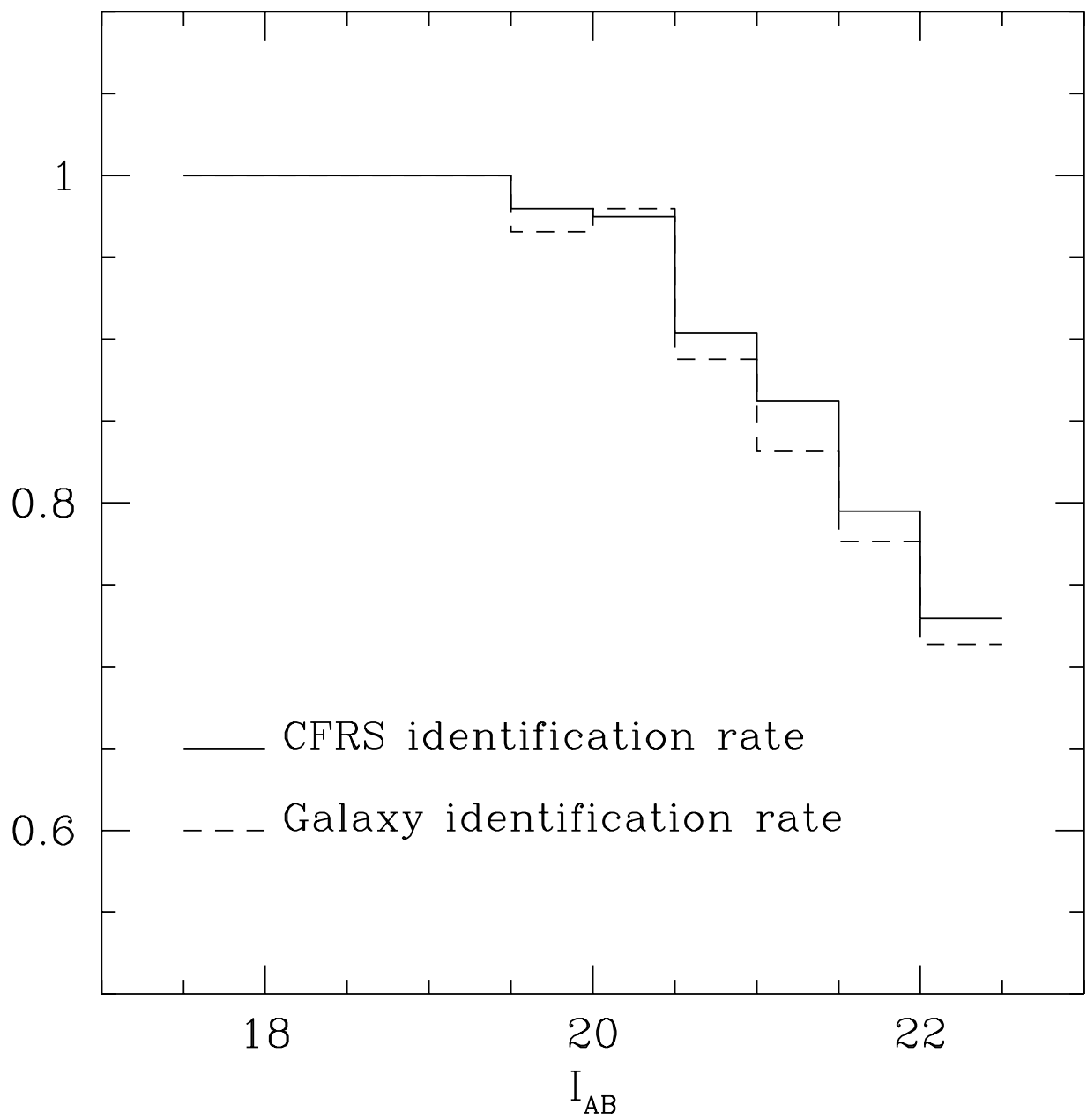

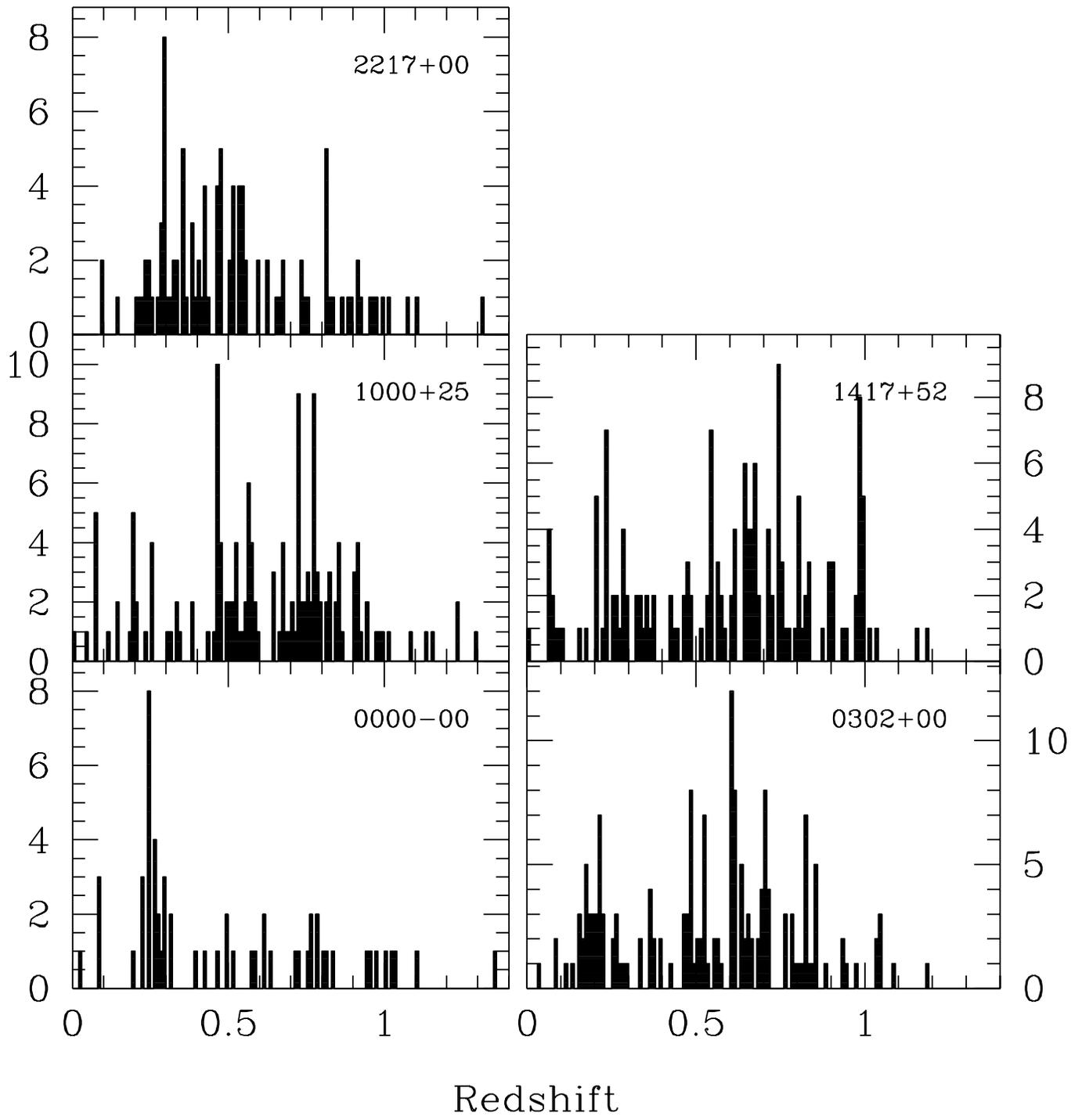

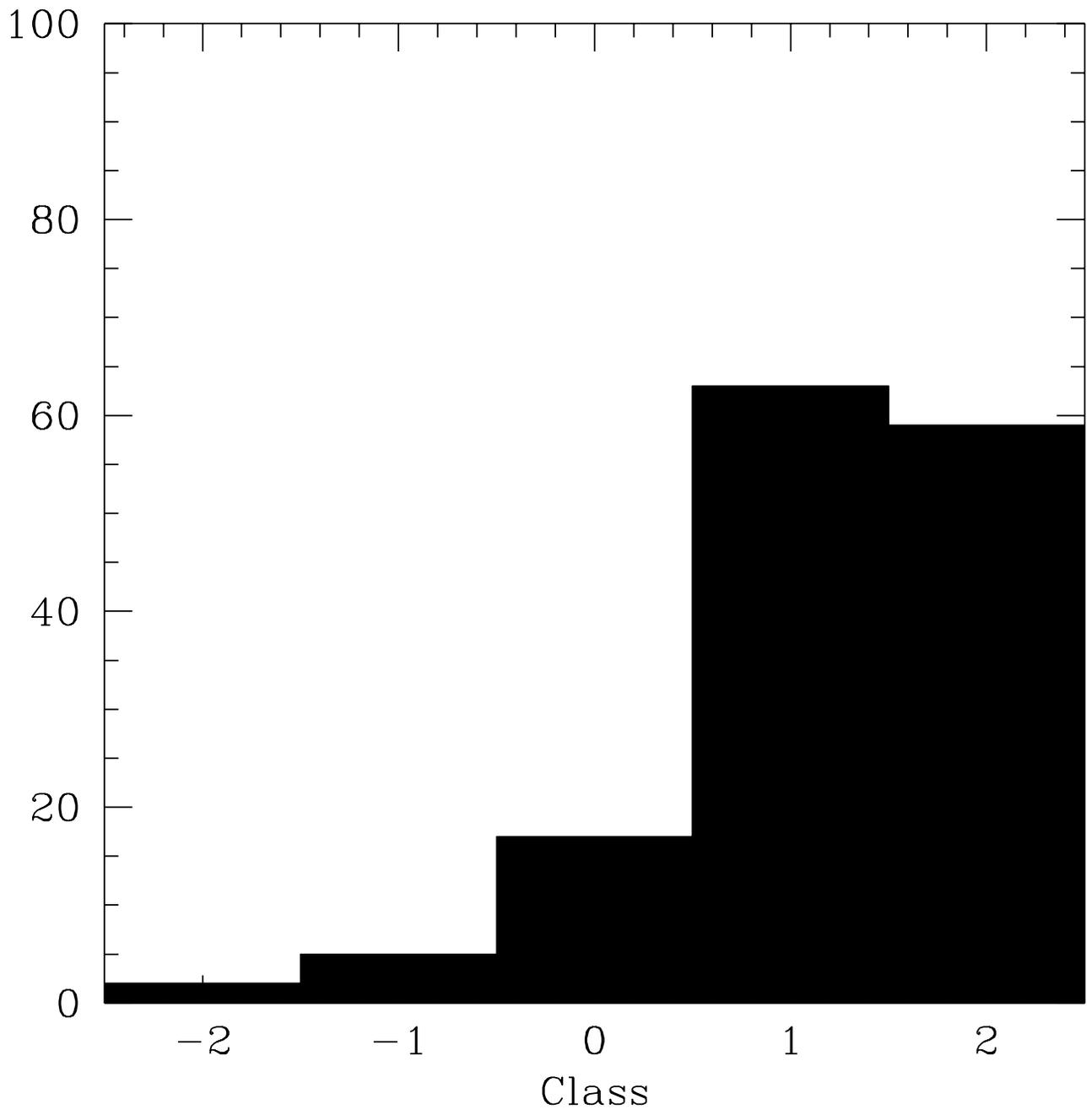

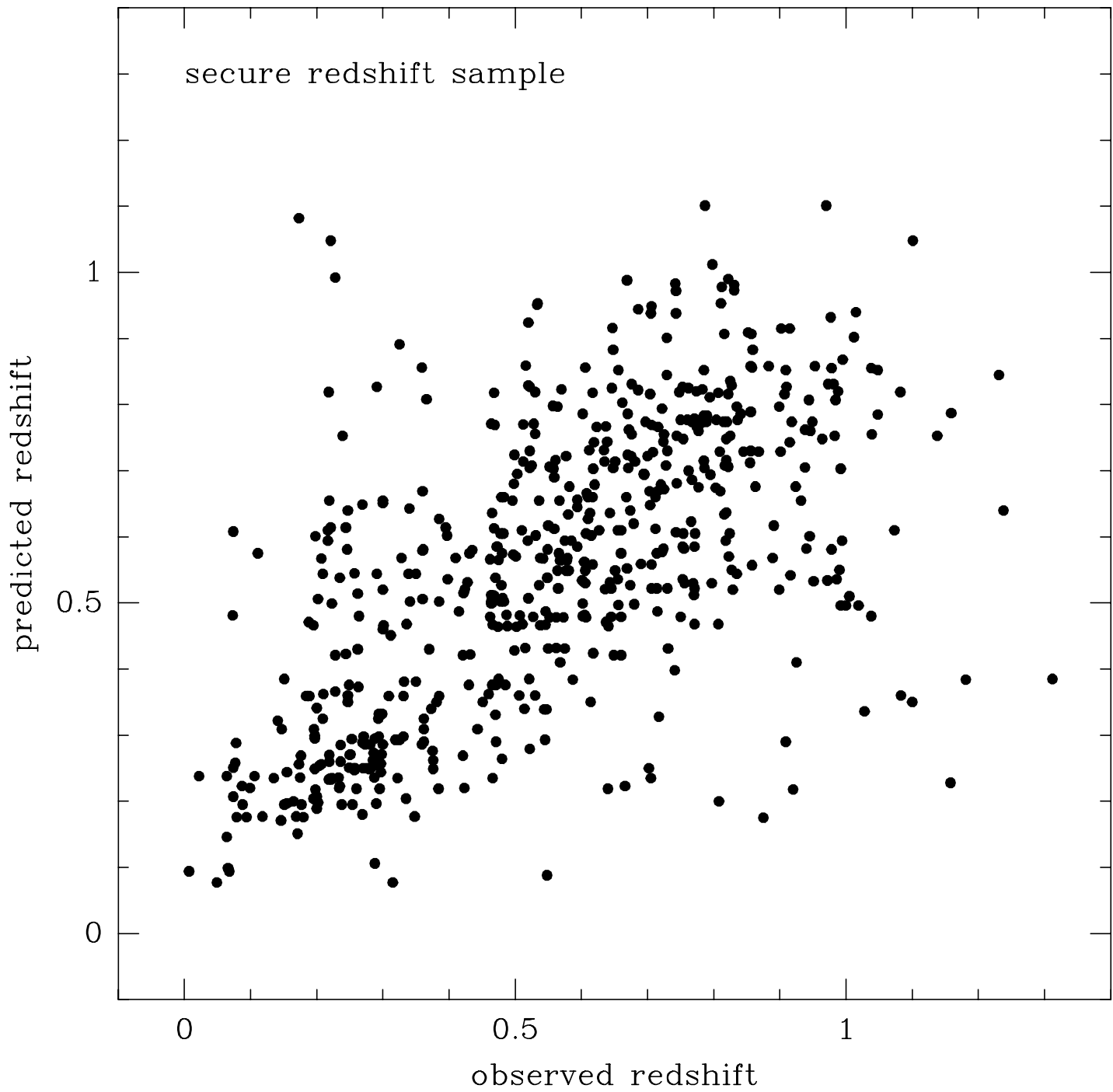

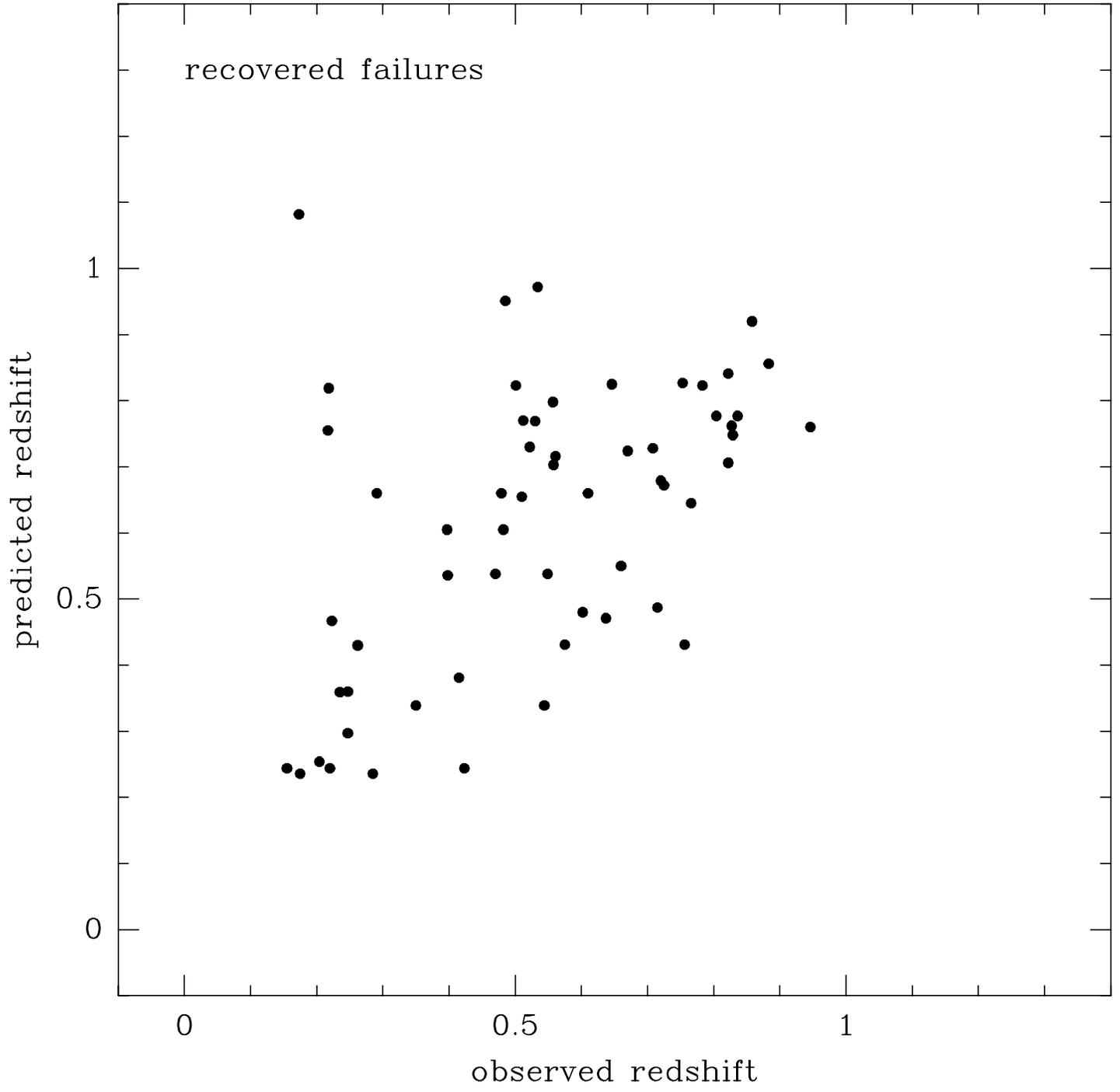

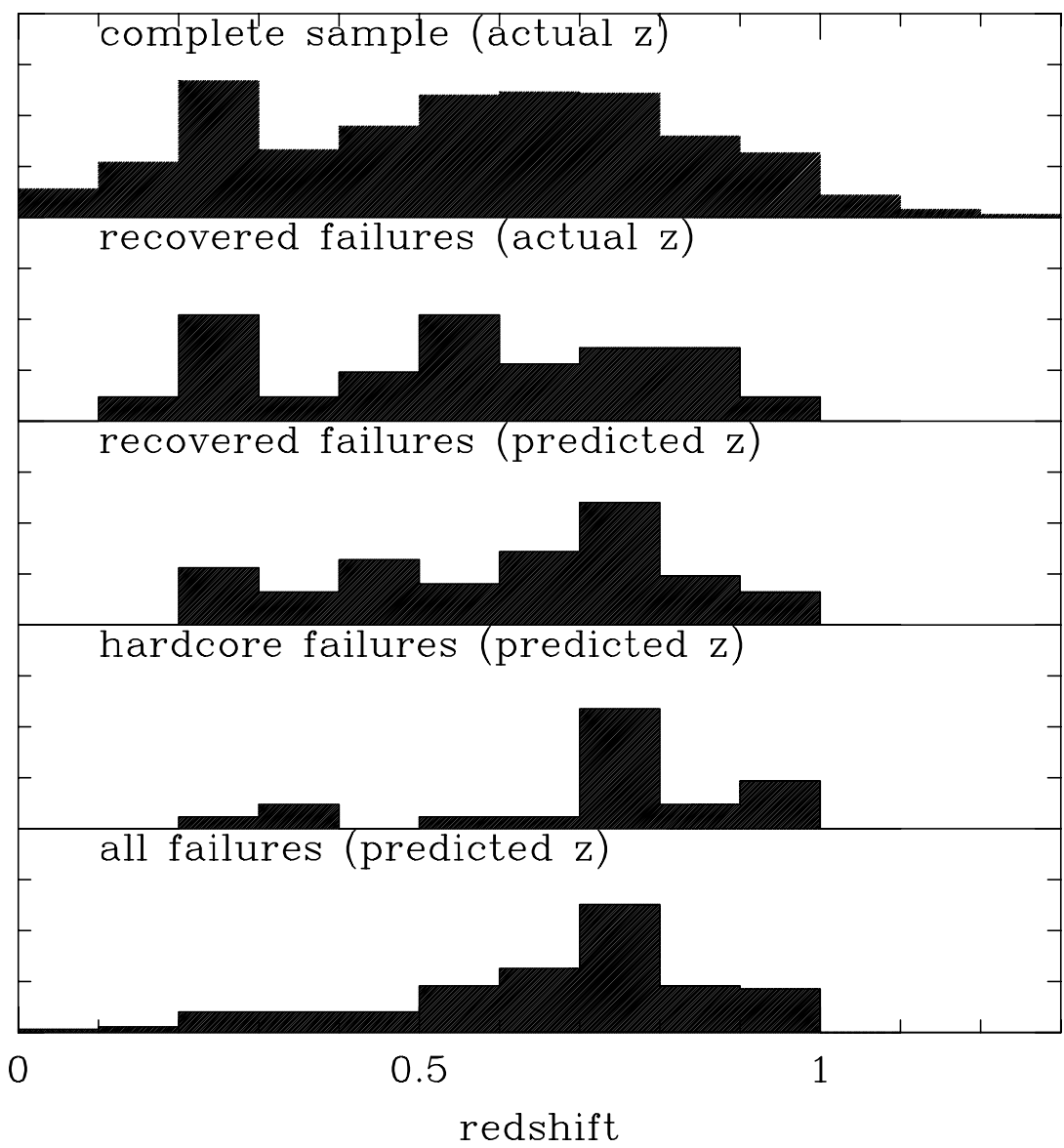

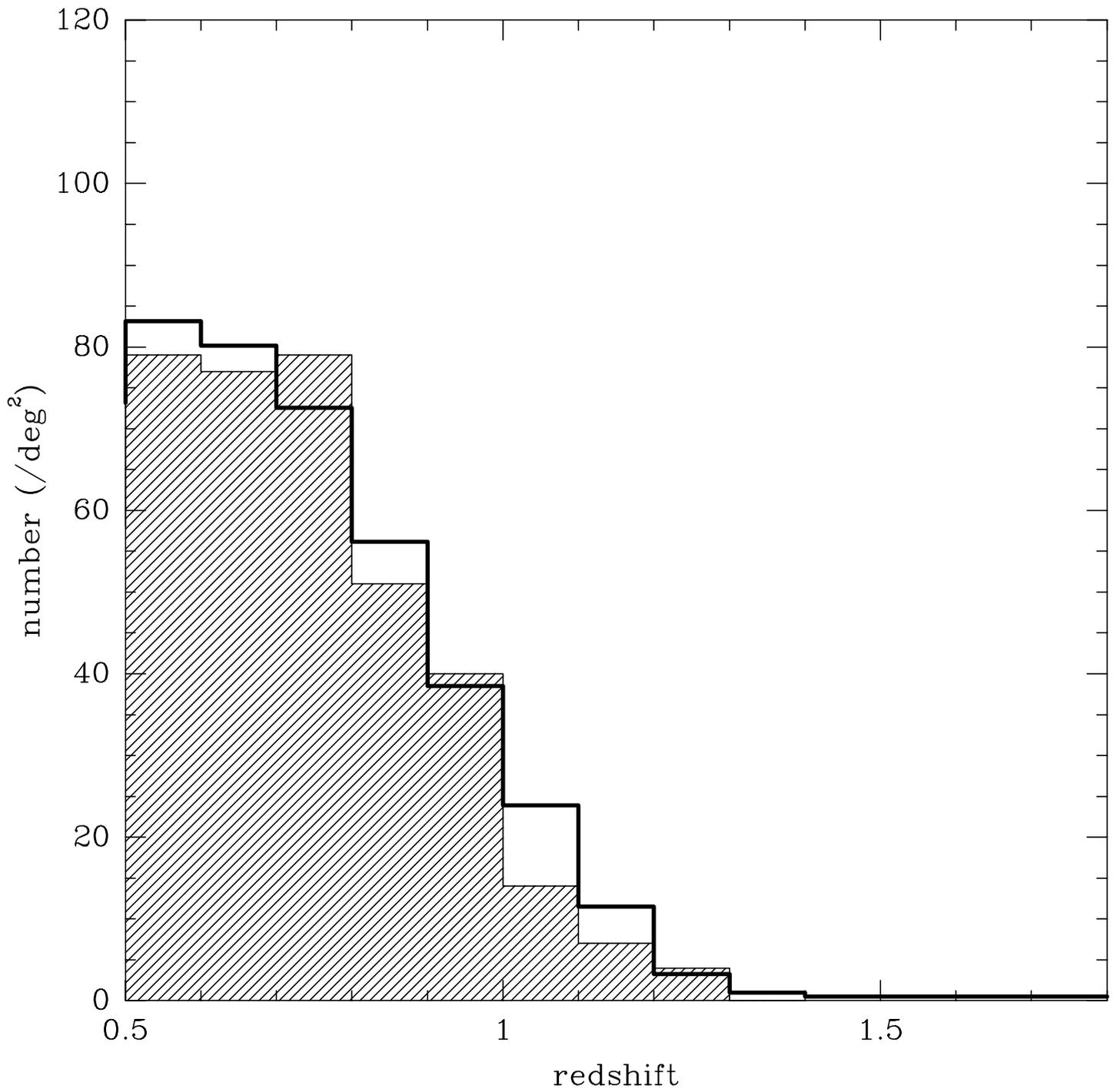